\documentclass[aps,twocolumn,floatfix,showpacs,superscriptaddress]{revtex4}

\pdfoutput=1

\usepackage{amsthm}
\usepackage[ruled]{algorithm}
\usepackage{algpseudocode}
\usepackage{slashbox}
\algnotext{EndFor}
\algnotext{EndIf}

\usepackage{graphicx}
\usepackage{subfig}
\usepackage{amsmath}
\usepackage{amssymb}
\usepackage{url}
\usepackage{mathrsfs}
\usepackage[dvipsnames]{xcolor}
\usepackage{hyperref}
\usepackage{enumitem}
\usepackage{natbib}
\usepackage{caption}
\hypersetup{colorlinks,%
citecolor=blue,%
filecolor=black,%
linkcolor=blue,%
urlcolor=blue
}

\newtheorem{defi}{Definition}
\newtheorem{prop}{Proposition}
\newtheorem{obs}{Observation}
\newtheorem{corol}{Corollary}

\begin{document}

\date{\today}








\title{Reconstructing the whole from its parts}

\begin{abstract}
The quantum marginal problem consists in deciding whether a given set of marginal reductions is compatible with the existence of a global quantum state or not. In this work, we formulate the problem from the perspective of dynamical systems theory and study its advantages with respect to the standard approach. The introduced formalism allows us to analytically determine global quantum states from a wide class of self-consistent marginal reductions in any multipartite scenario. In particular, we show that any self-consistent set of multipartite marginal reductions is compatible with the existence of a global quantum state, after passing through a depolarizing channel. This result reveals that the complexity associated to the marginal problem can be drastically reduced when restricting the attention to sufficiently mixed marginals. We also formulate the marginal problem in a compressed way, in the sense that the total number of scalar constraints is smaller than the one required by the standard approach. This fact suggests an exponential speedup in runtime when considering semi-definite programming techniques to solve it, in both classical and quantum algorithms. Finally, we reconstruct $n$-qubit quantum states from all the $\binom{n}{k}$ marginal reductions to $k$ parties, generated from randomly chosen mixed states. Numerical simulations reveal that the fraction of cases where we can find a global state equals 1 when $5\leq n\leq12$ and $\lfloor(n-1)/\sqrt{2}\rfloor\leq k\leq n-1$, where $\lfloor\cdot\rfloor$ denotes the floor function.
 \end{abstract}

\author{Daniel Uzcategui Contreras}
\affiliation{Departamento de F\'{i}sica, Facultad de Ciencias B\'{a}sicas, Universidad de Antofagasta, Casilla 170, Antofagasta, Chile}
\affiliation{Departamento de Física, Universidad de Concepción, 160-C Concepción, Chile}
\affiliation{Millennium Institute for Research in Optics, Universidad de Concepción, Concepción, Chile}

\author{Dardo Goyeneche}
\email{dardo.goyeneche@uantof.cl}
\affiliation{Departamento de F\'{i}sica, Facultad de Ciencias B\'{a}sicas, Universidad de Antofagasta, Casilla 170, Antofagasta, Chile}

\maketitle

\textit{Introduction.} The relation between the whole and its parts relies at the heart of quantum mechanics and quantum chemistry \cite{C63}. This fascinating topic, known as the quantum marginal problem (QMP), aims to answer the following question: \textit{given a set of multipartite quantum marginal reductions, is there a global quantum state compatible with them?} The QMP is closely related to the identification of separable quantum pure states in high dimensional bipartite systems \cite{YSWCNG21}, multipartite entanglement detection from nearest neighbour marginals \cite{PMMG18} and certification of quantum nonlocality from separable marginal reductions \cite{VLP14}. Furthermore, it is linked to the existence of absolutely maximally entangled (AME) states \cite{GALRZ15,HW_Table}, perfect tensors \cite{PYHP15} and quantum error correcting codes \cite{KL97}. Importantly, the study of the QMP in fermionic systems led to natural generalizations of the Pauli exclusion principle \cite{K06} and Hartree-Fock ansatz \cite{SBV17}, whereas for bosonic systems there are applications in quantum non-locality and self-testing of quantum states \cite{AFT21}. Furthermore, the QMP is closely related to the quantum channels compatibility problem \cite{HKMU21,GPS21}. Gentle introductions to the QMP and its applications can be found in the following PhD Thesis \cite{W14,S15,K17,W20,U22}.

Some partial solutions of the QMP are known. Higuchi, Sudbery and Szulc fully solved the case of 1-body marginals for $n$-qubit pure states \cite{HSS03}, whereas Franz solved the 3-qutrit case \cite{F02}.  Bravyi found necessary and sufficient conditions to solve the problem for  mixed states in 2-qubit systems \cite{B04}, which is equivalent to the $2\times2\times4$ case for pure states, see Chapter 2 in \cite{W14}. Later on, Klyachko found a set of necessary and sufficient conditions to solve the univariate quantum marginal problem for pure states, i.e. considering non-overlapping marginals and global pure states, for distinguishable particles \cite{K04} and fermions \cite{KA08}. On the other hand, generic 3-qubit pure states are completely determined by its two-qubit reductions \cite{LPW02} and generic pure quantum states are univocally reconstructed from reduced density matrices of a fraction of the parties \cite{LW02}. Moreover, the knowledge of all reductions to more than a half of the parties is sufficient to univocally reconstruct a pure quantum state, whereas all reductions to a half of the parties is not \cite{JL05}. These results reveal that \textit{the whole does not contain more information than its parts}, at least for the generic class of $n$-partite pure states. This last result also includes the generalized class of Dicke states \cite{PR09}. On the contrary, the generalized GHZ states for $n$-qubit systems are the only pure states that cannot be reconstructed from the knowledge of all its $n-1$-partite marginal reductions \cite{WL09}. A review about the existing partial solutions to the quantum marginal problem can be found here \cite{TV15}.\medskip

In this letter, we provide a partial answer to the QMP for an arbitrary number of parties and internal levels. First, we derive an analytical formula for the full set of hermitian operators having any prescribed set of self-consistent marginal reductions. Here, we show that a positive semi-definite solution can be found with positive probability provided that the marginals are compatible with the existence of a full-rank quantum state. Astonishingly, this probability seems to be equal to 1 in some multipartite scenarios, according to numerical simulations. We also analytically solve the QMP for any given set of sufficiently mixed marginal reductions. This reveals that an unexpected fraction of the quantum marginal problem can be analytically solved, extending some already existing results for quantum Markov chains \cite{HJPW04}. On the other hand, we reduce the number of scalar equations required to solve the QMP in multipartite scenarios, when the overlap in marginal reductions is sufficiently large. This fact might imply an exponential reduction in runtime for both classical and quantum algorithms.\medskip



We provide the proofs of our results in Supplemental Material A, whereas the analytical expressions (\ref{Qk1}) and (\ref{Qk2}) are demostrated in the Supplemental Material B. An iterative algorithm to reconstruct quantum states from marginal reductions is shown in Supplemental Material C. Here, the desired solutions can have any prescribed rank or spectrum, e.g. rank-one for finding pure states. We illustrate the usefulness of this algorithm for finding AME states.  In relatively low dimensions, this algorithm might be also used to induce that a given set of marginal reductions is not compatible with the existence of a quantum state. For instance, it does not converge for the case of pure states of 4 qubits with two-body maximally mixed reductions. However, in high dimensions it is hard to decide whether the algorithm has a slow convergence or it fails to converge. Indeed, it was not able to detect the AME state of 4 parties with 6 levels each, that actually exists \cite{RBBRLZ22}).
\medskip

\textit{Marginal imposition operator.}
 Consider a quantum system composed by $n$ parties, with any number of internal levels each, not necessarily equal. Let $\mathcal{I}$ be a set of letters of an alphabet with cardinality $|\mathcal{I}|=n$, used to denote parties of an entire quantum system. For instance, $\mathcal{I}=\{A,B,C\}$ applies to three partite systems. 
To simplify the notation, we assume that non-explicitly mentioned parties are associated to the maximally mixed state. For instance, $\sigma_{\mathcal{J}}$ means $\sigma_{\mathcal{J}}\otimes\frac{1}{|\mathcal{J}^c|}\mathbb{I}_{\mathcal{J}^c}$, except when $\sigma_{\mathcal{J}}$ is equal to a partial trace, and $\sigma_{\mathcal{J}}+\sigma_{\mathcal{J'}}$ means $\sigma_{\mathcal{J}}\otimes\frac{1}{|\mathcal{J}^c|}\mathbb{I}_{\mathcal{J}^c}+\sigma_{\mathcal{J'}}\otimes\frac{1}{|\mathcal{J'}^c|}\mathbb{I}_{\mathcal{J'}^c}$. Also, we restrict our attention to sets of marginal reductions that are compatible with the existence of a multipartite quantum state. To do that, we consider the notion of generator state.
\begin{defi}[Generator state]
A generator state is a multipartite quantum state $\sigma$ acting on $|\mathcal{I}|=n$ parties, that defines sets of $m$ marginal reductions by taking partial trace, i.e. $\sigma_{\mathcal{J}_j}=\mathrm{Tr}_{\mathcal{J}_j^c}[\sigma_{\mathcal{I}}]$, $j=1,\dots,m$.
\end{defi}
Note that marginal reductions arising from a generator state are always compatible with the existence of a quantum state. Now, we are in position to introduce the main tool of our work, the \textit{marginal imposition operator}.
\begin{defi}[Marginal Imposition Operator]
Let $\sigma_{\mathcal{J}}$ be a $|\mathcal{J}|$-partite quantum state, for $\mathcal{J}\subset\mathcal{I}$. The marginal imposition operator is defined as follows:
\begin{equation}\label{qmo}
\mathcal{Q}_{\sigma_{\mathcal{J}}}(\rho_{\mathcal{I}}):=\rho_{\mathcal{I}}-\rho_{\mathcal{J}}+\sigma_{\mathcal{J}}.
\end{equation}
\end{defi}
Note that the trace preserving map $\mathcal{Q}_{\sigma_{\mathcal{J}}}$ takes a quantum state $\rho_{\mathcal{I}}$, removes all the information contained in the subset of parties $\mathcal{J}$, and imposes the marginal reduction $\sigma_{\mathcal{J}}$.
 
Let us show two fundamental properties of the operator (\ref{qmo}): it imposes any given marginal reduction $\sigma_{\mathcal{J}}$, and it does not perturb the information stored in the complementary subset of parties $\mathcal{J}^c:=\mathcal{I}\setminus\mathcal{J}$.
\begin{prop}[Imposition]\label{impo_marginals}
Let $\sigma_{\mathcal{J}}$ be a quantum state, with $\mathcal{J}\subset\mathcal{I}$. Therefore, $\mathrm{Tr}_{\mathcal{J}^c}[\mathcal{Q}_{\sigma_{\mathcal{J}}}(\rho_{\mathcal{I}})]=\sigma_{\mathcal{J}}$, for any $\rho_{\mathcal{I}},\sigma_{\mathcal{J}}$. Also, if $\mathcal{K}\subset\mathcal{I}$ such that $\mathcal{J}\cap\mathcal{K}=\emptyset$, then 
$\mathrm{Tr}_{\mathcal{K}^c}[\mathcal{Q}_{\sigma_{\mathcal{J}}}(\rho_{\mathcal{I}})]=\mathrm{Tr}_{\mathcal{K}^c}[\rho_{\mathcal{I}}]=\rho_{\mathcal{K}}$, for any $\rho_{\mathcal{I}},\sigma_{\mathcal{J}}$.
\end{prop}
To simplify the notation, from now on we omit the subindex $\mathcal{I}$ in the state $\rho_{\mathcal{I}}$ and denote the operator $\mathcal{Q}_{\sigma_{\mathcal{J}}}$ as $\mathcal{Q}_{\mathcal{J}}$. Let us show that operator (\ref{qmo}) is idempotent.
\begin{prop}[Idempotence]\label{orthproj}
The marginal imposition operator $\mathcal{Q}_{\mathcal{J}}(\rho)$ is idempotent. That is, it satisfies the relation $\mathcal{Q}_{\mathcal{J}}\circ\mathcal{Q}_{\mathcal{J}}(\rho)=\mathcal{Q}_{\mathcal{J}}(\rho)$.
\end{prop}

Proposition \ref{orthproj} tells us that imposing once or twice a marginal reduction $\sigma_{\mathcal{J}}$ through operator (\ref{qmo}) are equivalent actions, in agreement with Proposition \ref{impo_marginals}. When considering sets of $m>1$ marginal reductions $\sigma_{\mathcal{J}_1},\dots,\sigma_{\mathcal{J}_m}$, the composite marginal imposition operator is a natural generalization of (\ref{qmo}). That is,
\begin{equation}\label{qmo_comp}
\mathcal{Q}_{\mathcal{J}_1,\dots,\mathcal{J}_m}=\mathcal{Q}_{\mathcal{J}_m}\circ\dots\circ\mathcal{Q}_{\mathcal{J}_1}.
\end{equation}
From the operational point of view, it is convenient to find (\ref{qmo_comp}) as an explicit expression of the marginal reductions. To do that, it is convenient to introduce first some further definitions. Let $\mathscr{J}$ be the set formed by all possible intersections of the subsets $\mathcal{J}_1,\dots,\mathcal{J}_m\in\mathcal{I}$. That is,
\begin{widetext}
\begin{equation}\label{J}
\mathscr{J}=\{\mathcal{J}_1,\dots,\mathcal{J}_m,\mathcal{J}_1\cap\mathcal{J}_2,\dots,\mathcal{J}_{m-1}\cap\mathcal{J}_m,\mathcal{J}_{1}\cap\mathcal{J}_2\cap\mathcal{J}_3,\dots,\mathcal{J}_{m-2}\cap\mathcal{J}_{m-1}\cap\mathcal{J}_m,\dots\}.
\end{equation}
\end{widetext}
When all the sets $\mathcal{J}_i$ do intersect, $i=1,\dots,m$, then it is easy to see that $|\mathscr{J}|=\sum_{i=1}^{m}\binom{m}{i}=2^m-2$. For instance, if $m=3$, $\mathcal{J}_1=\{A,B\}$, $\mathcal{J}_2=\{A,C\}$ and $\mathcal{J}_3=\{B,C\}$, then $\mathscr{J}=\{A,B,C,AB,AC,BC\}$ and $|\mathscr{J}|=6$. Also, disjoint subsets $\mathcal{J}_1,\dots,\mathcal{J}_m$ imply that $|\mathscr{J}|=m$. Additionally, let $\nu(X)$ be the number of sets $\mathcal{J}_i$ involved in the implicit expression of a set $X\in\mathscr{J}$. For instance, $\nu(\mathcal{J}_i)=1$ for any $i\in\{1,\dots,n\}$, $\nu(\mathcal{J}_i\cap\mathcal{J}_j)=2$, for any $i<j$, $\nu(\mathcal{J}_i\cap\mathcal{J}_j\cap\mathcal{J}_k)=3$, for any $i<j<k$, and so on. Note that $\nu(X)$ is well defined when $X$ is implicitly defined as a function of intersections of sets $\mathcal{J}_i$, whereas it \textit{is not a function} when associating explicit parties to $\mathcal{J}_i$. For instance, if $\mathcal{J}_1=\{A,B\}$, $\mathcal{J}_2=\{B,C\}$ and $\mathcal{J}_3=\{B\}$, then $2=\nu(\mathcal{J}_1\cap\mathcal{J}_2)=\nu(\mathcal{J}_3)=1!$  By using (\ref{qmo}) and the above introduced notation, we obtain a closed form of (\ref{qmo_comp}).
\begin{prop}[Closed form]\label{prop:closed}
The composite marginal imposition operator (\ref{qmo_comp}) is given by
\begin{equation}\label{qmo2}
\mathcal{Q}_{\mathcal{J}_1,\dots,\mathcal{J}_m}(\rho)=\rho+\sum_{X\in\mathscr{J}}(-1)^{\nu(X)}(\rho_X-\sigma_X).
\end{equation}
\end{prop}
For the three-partite case, an operator having similar properties than (\ref{qmo2}) was used to study global quantum correlations based on bipartite marginal reductions \cite{CGMP14}. Also, for the special case of non-overlapped marginals, the closed form (\ref{qmo2}) reduces to
\begin{equation}\label{qmo_nonoverlapping}
\mathcal{Q}_{\mathcal{J}_1,\dots,\mathcal{J}_m}(\rho)=\rho-(\rho_{\mathcal{J}_1}+\dots+\rho_{\mathcal{J}_m})+(\sigma_{\mathcal{J}_1}+\dots+\sigma_{\mathcal{J}_m}).
\end{equation}
Due to linearity of the partial trace, the normalized convex combination of quantum states having identical marginal reductions has the same marginal reductions. This fact is reflected in the following result.
\begin{prop}[Convexity]\label{Qconvex}
Let $\rho$ and $\eta$ be any two multipartite quantum states acting on the same composite Hilbert space. Therefore, \begin{equation}
\mathcal{Q}(\lambda\rho+(1-\lambda)\eta)=\lambda\mathcal{Q}(\rho)+(1-\lambda)\mathcal{Q}(\eta),
\end{equation} 
for any $0\leq\lambda\leq1$, and any subsets of parties $\mathcal{J}_1,\dots,\mathcal{J}_m\subset\mathcal{I}$, not necessarily disjoined.  
\end{prop}
A necessary condition to reconstruct states from marginal reductions is the self-consistency of marginal reductions, introduced below.
\begin{defi}[self-consistency]\label{compatible}
Two marginal reductions $\sigma_{\mathcal{J}_1}$ and $\sigma_{\mathcal{J}_2}$ are self-consistent if $\mathrm{Tr}_{\tilde{\mathcal{J}}_2^{c}}(\sigma_{\mathcal{J}_1})=\mathrm{Tr}_{\tilde{\mathcal{J}_1}^{c}}(\sigma_{\mathcal{J}_2})=:\sigma_{\mathcal{J}_1\cap\mathcal{J}_2}$. A set of $m$ marginal reductions $\sigma_{\mathcal{J}_1},\dots,\sigma_{\mathcal{J}_m}$ are self-consistent if they are pairwise self-consistent.
\end{defi}
For instance, if $\mathcal{J}_1=\{A,B\}$ and $\mathcal{J}_2=\{B,C\}$, the self-consistency condition implies that $\mathrm{Tr}_A(\sigma_{AB})=\mathrm{Tr}_C(\sigma_{BC})=\sigma_B$. Now, let us show that self-consistency of marginal reductions is closely related to commutativity of marginal imposition operators.
\begin{prop}[self-consistency \& conmutativity]\label{Qcommute}
Two marginal imposition operators $\mathcal{Q}_{\mathcal{J}}$ and $\mathcal{Q}_{\mathcal{J}'}$ do commute if and only if the imposed marginals, $\sigma_{\mathcal{J}}$ and $\sigma_{\mathcal{J}'}$ respectively, are self-consistent. \end{prop}
In particular, note that any pair of non-overlapping marginal reductions are self-consistent and their related imposition operators do commute. 

From Prop. (\ref{orthproj}), a state $\rho$ that already contains some given marginals is invariant under the action of the marginal imposition operator. That is, such states are fixed points of the operator (\ref{qmo_comp}).
\begin{prop}[Fixed points]\label{prop:fixed}
An hermitian operator $\rho$ satisfies   $\mathrm{Tr}_{\mathcal{J}^c_1}(\rho)=\sigma_{\mathcal{J}_1},\dots,\mathrm{Tr}_{\mathcal{J}^c_m}(\rho)=\sigma_{\mathcal{J}_m}$ if and only if $\rho$ is a fixed point of the composite marginal imposition operator $\mathcal{Q}_{\mathcal{J}_1,\dots,\mathcal{J}_m}$.
\end{prop}
Note that the entire set of fixed points of $\mathcal{Q}_{\mathcal{J}_1,\dots,\mathcal{J}_m}$ is not necessarily a subset of the positive semi-definite cone. Indeed, this set intersects the positive semi-definite cone when there are solutions to the QMP problem. In particular, there is a tangential intersection when there is a unique solution. \medskip

\textit{Main results.} We start by observing that the closed form (\ref{qmo2}) allows us to encode the QMP in a single linear equation, complemented by the normalization and positivity conditions. That is,
\begin{prop}\label{prop:condi} A set of self-consistent marginal reductions $\sigma_{\mathcal{J}_1},\dots,\sigma_{\mathcal{J}_m}$ is compatible with the existence of a quantum state $\rho$ if and only if
\begin{eqnarray}\label{qmp}
\sum_{X\in\mathscr{J}}(-1)^{\nu(X)}(\rho_X-\sigma_X)=0,\nonumber\\
\mathrm{Tr}(\rho)=1,\quad\mbox{and}\quad\rho\geq0.
\end{eqnarray}
\end{prop}
In some multipartite scenarios, Proposition \ref{prop:condi} implies a \textit{strong reduction in the number of linear equations required to solve the QMP}. For instance, the standard approach to solve the QMP implies to consider $\binom{n}{k}$ marginals, when considering all reductions to $k$ bodies. Thus, in the concrete scenario of $n$ qubit systems with all $k$-body reductions, there are $2^{2k}\binom{n}{k}$ scalar equations, arising from the Bloch decomposition of all the marginals. On the other hand, Prop. \ref{prop:condi} implies $2^{2n}$ scalar equations, regardless on the value of $k$. Note that there is an exponential advantage of our approach for large values of $k$: e.g. if $0.7n<k<0.9n$, then $2^{2k}\binom{n}{k}-2^{2n}>2^{1.4n}$, for any $n>5$. The advantage is due to a reduction in the number of redundant equations, when considering a large number of overlapped marginals. Actually, if the overlap is not-so-strong (i.e. if $k<0.5n$) then our approach does not have any advantage with respect to the standard formulation, for $n$-qubit systems. This situation can be straightforwardly extrapolated to any multipartite scenario, with any number and size of reductions.

The above described compressed way to pose the QMP might have interesting consequences in the runtime of its numerical resolution by considering SDP techniques. To show this, let us now briefly mention the cost of solving SDP problems with some currently existing algorithms. Firstly, we mention that Khachiyan \cite{khachiyan1979polynomial} proved that Linear Programming can be solved in a polynomial number of operations, i.e. $O(N^3(N+m)L)$, where $N$ is the order of the matrix, $m$ the number of scalar constraints and $L$ is an input size parameter. Later, Karmarkar designed an algorithm with complexity bounded by $O(m^{3/2}N^2L)$ operations \cite{karmarkar1984new}, whereas Alizadeh found an algorithm based on interior point methods with a runtime $\tilde{O}(\sqrt{m}(m+N^3)L)$ \cite{alizadeh1995interior,nesterov1994interior}. Here, $\tilde{O}$ means that $\mathrm{polylog}$ factors are suppressed. A more efficient technique based on the interior point method, found by Jiang \textit{et. al.} \cite{jiang2020faster}, has runtime $\tilde{O}\bigl(\sqrt{N}(mN^2+m^{\omega}+N^{\omega})\log(1/\epsilon)\bigr)$, where $\omega$ is the exponent of matrix multiplication and $\epsilon$ is the desired accuracy. On the other hand, some algorithms were designed to solve SDP problems on quantum computers, having a remarkable speedup in relatively high dimensions. For instance, Brandao and Svore presented a quantum algorithm with worst-case running time $\tilde{O}(N^{1/2}m^{1/2}s^2)$, where $s$ is the row-sparcity of the input matrix \cite{brandao2017quantum}. Also, another quantum algorithm, implementable in NISQ computers, has been recently developed by Bharti \textit{et. al.} \cite{bharti2021nisq}. In all these classical and quantum algorithms, the runtime is an increasing function on the number of constraints, growing at least like $O(m^{1/2})$. As a consequence of this fact, the constraints shown in Proposition \ref{prop:condi} might imply an \textit{exponential speed up} for reconstructing quantum states from its marginals when considering currently existing SDP techniques, in a wide range of scenarios. We warn the reader that this exponential speed up would not imply a reduction in the complexity class of the quantum marginal problem, which is actually QMA-Complete \cite{L06}.\medskip

Even assuming above potential speed up, the QMP is still hopeless in very high scenarios. In what follows, we show an analytical way to solve this problem for a wide range of choices of marginal reductions. That is, we provide a method that does not require to solve any set of equations nor implementing any multi-step algorithm, allowing us to  scale to very large scenarios. \begin{prop}\label{prop:Q_state}
Let $\sigma_{\mathcal{J}_1},\dots,\sigma_{\mathcal{J}_m}$ be a set of self-consistent marginal reductions. If $\mathcal{Q}_{\mathcal{J}_1,\dots,\mathcal{J}_m}(\rho)\geq0$, for a given quantum state $\rho$, then $\mathcal{Q}_{\mathcal{J}_1,\dots,\mathcal{J}_m}(\rho)$ is a quantum state containing all the above reductions. 
\end{prop}
Prop. \ref{prop:Q_state} is a straight consequence of Props. \ref{impo_marginals} and \ref{Qcommute}. At this point, one might wonder how frequently the relation $\mathcal{Q}_{\mathcal{J}_1,\dots,\mathcal{J}_m}(\rho)\geq0$ holds. Below, we show that there is a  set of states $\rho$, having positive volume in the Hilbert space, such that $\mathcal{Q}_{\mathcal{J}_1,\dots,\mathcal{J}_m}(\rho)\geq0$, provided that a full-rank generator state $\sigma$ exists for the involved marginals. In other words, there is a positive probability to find a suitable state $\rho$ at random when marginal reductions are compatible with the existence of a full-rank quantum state.
\begin{prop}\label{prop:positiveV}
Let $\sigma_{\mathcal{J}_1},\dots,\sigma_{\mathcal{J}_m}$ be a set of marginal reductions compatible with the existence of a full-rank quantum state $\sigma$. Therefore, there is a set $\Gamma$ of quantum states, having a positive volume in the space of density matrices, such that $\mathcal{Q}_{\mathcal{J}_1,\dots,\mathcal{J}_m}(\rho)\geq0$, for any $\rho\in\Gamma$.
\end{prop}
Moreover, numerical simulations indicate that the maximally mixed state $\rho=\mathbb{I}/d$, is a suitable choice in a wide range of $d$-dimensional scenarios. Table 1 in Supplemental Material D shows the probability of obtaining a quantum state $\mathcal{Q}_{\mathcal{J}_1,\dots,\mathcal{J}_m}(\mathbb{I}/d)>0$, where $\sigma$ is a generator state chosen at random according to the Hilbert-Schmidt measure. This study is implemented in systems composed from 3 to 12 qubits. Astonishingly, our method seems to reconstruct a quantum state from marginals defined through any randomly chosen generator state $\sigma$, in a wide range of scenarios. Here, random states are taken according to the Hilbert-Schmidt measure.
\medskip


\textit{Solving a fraction of the QMP.} The quantum marginal problem belongs to the complexity class QMA \cite{L06}, as well as its fermionic version \cite{LCV07}. Nonetheless, we show below that a fraction of this problem, corresponding to the case where the marginal reductions are sufficiently mixed, can be analytically solved in any multipartite scenario. Before showing the result, we recall the definition of depolarizing quantum channel, $\Delta_{\epsilon}(\rho)=(1-\epsilon)\rho+\epsilon\mathbb{I}/d$.
\begin{prop}\label{prop:verymixed}
Let $\sigma_{\mathcal{J}_1},\dots,\sigma_{\mathcal{J}_m}$ be any given set of self-consistent marginal reductions, according to Proposition \ref{Qcommute}. Therefore, the marginals ${\tilde{\sigma}_{\mathcal{J}_j}=\Delta_{\epsilon}(\sigma_{\mathcal{J}_j})}$, $j=1,\dots,m$, are compatible with the existence of a global quantum state, provided that ${|\mathscr{J}|(1-\epsilon)(d-1)\leq1}$, where $\mathscr{J}$ is defined in (\ref{J}).
\end{prop}
Note that quantum states sufficiently close to the maximally mixed state are fully separable for bipartite \cite{HHH96} and multipartite quantum systems  \cite{GB03}. Here, an interesting question arises: \textit{are the global states provided by Prop. \ref{prop:verymixed} fully separable?} This is a likely situation, as there is a close relation between separability and the quantum marginal problem \cite{YSWCNG21}. Indeed, the ball containing all reconstructed quantum states shown in Prop. \ref{prop:verymixed} for $n$ partite quantum systems, centered in the maximally mixed states, has radius $r_{rec}=\frac{1}{(d-1)|\mathscr{J}|}$. This value is much smaller than the radius of a ball composed by fully separable quantum states, given by $r_{sep}=2^{-(n/2-1)}$ , with $n>2$ \cite{GB03}. That is, for the smallest local dimension (qubits) and the smallest possible cardinality $|\mathscr{J}|=2$, i.e. when considering two non-overlapped reductions, we have that $r_{rec}\approx2^{-(n+1)}<2^{-(n/2-1)}= r_{sep}$.\medskip

There are certain classes of quantum states, called quantum Markov chains, for which a \textit{universal recovery map} exists \cite{HJPW04}. That is, there is a physical procedure to reconstruct a quantum state having a desired marginal reduction. In the three partite case, short quantum Markov chains are those quantum states having a vanishing conditional mutual information,  $I(A:C|B)=0$. For instance, the following CPTP map does the job: $(\mathcal{I}_A\otimes\tau_{B\rightarrow BC})(\rho_{AB})=\rho_{ABC}$, where $\tau_{B\rightarrow BC}(X_B)=\rho_{BC}^{1/2}(\rho_N^{-1/2}X_B\rho_N^{-1/2}\otimes\mathbb{I}_C)\rho_{BC}^{1/2}$ is the Petz map \cite{P86} and the short quantum Markov chain $\rho_{ABC}$ contains the marginal reduction $\rho_{AB}$. Similar results hold for approximate quantum Markov chains \cite{SFR16}. Relating these facts with our results, note that for three partite systems, with $m=1$ and $\mathcal{J}_1=\{AC\}$, all states reconstructed by Prop. \ref{prop:verymixed} are short quantum Markov chains, see Corollary 7 in \cite{HJPW04}. Here, we stress that the reconstruction provided by operator (\ref{qmo_comp}) goes beyond exact or approximate quantum Markov chains. Indeed, if the generator $\sigma$ is a genuinely entangled full-rank quantum state then $\mathcal{Q}_{\mathcal{J}_1,\dots,\mathcal{J}_m}(\rho)\geq0$, for a set of states $\rho$ that has a positive measure in the Hilbert space, see Prop. \ref{prop:positiveV}. In other words, operator (\ref{qmo_comp}) allows us to recover multipartite entangled quantum states having full-rank, with positive probability, from choosing $\rho$ at random. 

Prop. \ref{prop:verymixed} also shows that $\mathcal{Q}_{\mathcal{J}_1,\dots,\mathcal{J}_m}\circ\Delta_{\epsilon}$ is a positive map, for $\epsilon$ sufficiently close to 1, despite $\mathcal{Q}_{\mathcal{J}_1,\dots,\mathcal{J}_m}$ is not. However, it is simple to show that this composite map is not completely positive, for any $\epsilon>0$ and any multipartite scenario. Note that the so-called \textit{universal state inversion} map is also positive but not completely positive \cite{RBCHM01,H05}. The imposibility to physically implement the map (\ref{qmo_comp}) might be closely related to the facts that quantum states can neither be cloned \cite{wootters1982single} nor deleted \cite{PB00}.\medskip

\textit{Analytical expressions.} Some analytical expressions of the composite operator (\ref{qmo2}) can be derived for any number of parties $n$ and internal levels. When considering all single particle reductions $\mathcal{J}_1$ to $\mathcal{J}_{n}$, with $|\mathcal{J}_i|=1$, $i=1,\dots,n$, we have 
\begin{equation}\label{Qk1}
   \mathcal{Q}_{\mathcal{J}_n} \circ \ldots \circ \mathcal{Q}_{\mathcal{J}_1}(\mathbb{I})= \underset{ i \in \mathcal{I} }{ \sum } \sigma_i -(n-1)\mathbb{I}.
\end{equation}
whereas for all two-body marginals  $\sigma_{\mathcal{J}_1}$ to $\sigma_{\mathcal{J}_{\binom{n}{2}}}$, with $|\mathcal{J}_i|=2$, $i=1,\dots,\binom{n}{2}$, we have
\begin{equation}\label{Qk2}
   \mathcal{Q}_{\mathcal{J}_{\binom{n}{2}}} \circ \ldots \circ \mathcal{Q}_{\mathcal{J}_1}(\mathbb{I})= \underset{  j \in \mathcal{J}  }{ \sum } \sigma_j - (n-2)\underset{ i \in \mathcal{I} }{ \sum } \sigma_i +\xi\,\mathbb{I},
\end{equation}
with $\xi=1+\frac{n^2}{2}-\frac{3}{2}n$. The proofs for results (\ref{Qk1}) and (\ref{Qk2}) are shown in Supplemental Material B.
\medskip

\textit{Conclusions.} We studied the quantum marginal problem from the perspective of dynamical systems theory. Solutions to this problem are one-to-one related to fixed points of the so-called marginal imposition operator, see Definition \ref{qmo}. This approach allowed us to encode the multipartite quantum marginal problem in a single linear equation in the full space complemented by positivity and normalization, regardless on the number of parties, internal levels and reductions considered, see Proposition \ref{prop:condi}. This compressed formulation revealed an exponential reduction in the number of scalar equations required to defines the quantum marginal problem. For instance, for  $n$-qubit systems and all reductions to $k$ qubits, an exponential reduction occurs when $0.7n<k<0.9n$. This fact suggests an exponential speedup when solving the quantum marginal problem with SDP techniques, in both classical and quantum algorithms. Furthermore, we demonstrated that any given set of self-consistent marginal reductions is compatible with the existence of a global quantum state, after passing through a depolarizing channel, see Proposition \ref{prop:verymixed}. As a further result, we designed an iterative algorithm to study the quantum state marginal problem from any given set of overlapped quantum marginals in any multipartite scenario, see Supplemental Material C. 

Finally, we provide an analytical formula for reconstruction of quantum states from its marginal reductions that produces a solution with positive probability, as long as the marginal reductions are compatible with the existence of a full-rank global state. Moreover, this probability seems to be equal to one in some scenarios, see Supplemental Material D.\\

\textit{Acknowledgements.} It is a pleasure to thank Lin Chen, Yi Shen, Felix Huber and Gabriel Senno for valuable comments. DG and DU are supported by Grant FONDECYT Iniciaci\'{o}n number 11180474, Chile. DU also acknowledges support from Project ANT1956, Universidad de Antofagasta, Chile, and ANID – Millennium Science Initiative Program – ICN17\_012.
\bigskip

\onecolumngrid

\appendix

\subsection*{Supplemental Material A: Proof of results}
\noindent\textbf{Proposition 1}
\textit{Let $\sigma_{\mathcal{J}}$ be a quantum state, with $\mathcal{J}\subset\mathcal{I}$. Therefore, $\mathrm{Tr}_{\mathcal{J}^c}[\mathcal{Q}_{\sigma_{\mathcal{J}}}(\rho)]=\sigma_{\mathcal{J}}$. Also, if $\mathcal{K}\subset\mathcal{I}$ such that $\mathcal{J}\cap\mathcal{K}=\emptyset$, then 
$\mathrm{Tr}_{\mathcal{K}^c}[\mathcal{Q}_{\sigma_{\mathcal{J}}}(\rho)]=\mathrm{Tr}_{\mathcal{K}^c}[\rho]=\rho_{\mathcal{K}}$.}
\begin{proof}
From (1), main text, we have that
\begin{eqnarray}\label{imposed_marginal}
\mathrm{Tr}_{\mathcal{J}^c}\left[Q_{\sigma_{\mathcal{J}}}(\rho) \right] &=& \mathrm{Tr}_{\mathcal{J}^c}\left[\rho - \rho_{\mathcal{J}}+\sigma_{\mathcal{J}} \right]\nonumber\\
&=& \mathrm{Tr}_{\mathcal{J}^c}\left[\rho \right]- \mathrm{Tr}_{\mathcal{J}^c}\left[\rho_{\mathcal{J}}\right]+\mathrm{Tr}_{\mathcal{J}^c}\left[\sigma_{\mathcal{J}} \right]\nonumber\\
&=& \mathrm{Tr}_{\mathcal{J}^c}\left[\sigma_{\mathcal{J}} \right]\nonumber\\
&=&\sigma_{\mathcal{J}}.
\end{eqnarray}
Similarly,
\begin{eqnarray}\label{imposed_marginal_k}
\mathrm{Tr}_{\mathcal{K}^c}\left[Q_{\sigma_{\mathcal{J}}}(\rho) \right]
&=& \mathrm{Tr}_{\mathcal{K}^c}\left[\rho - \rho_{\mathcal{J}}+\sigma_{\mathcal{J}}\right]\nonumber\\
&=& \mathrm{Tr}_{\mathcal{K}^c}\left[\rho \right]- \mathrm{Tr}_{\mathcal{K}^c}\left[\rho_{\mathcal{J}}\right]+\mathrm{Tr}_{\mathcal{K}^c}\left[\sigma_{\mathcal{J}} \right]\nonumber\\
&=&\mathrm{Tr}_{\mathcal{K}^c}\left[\rho \right]\nonumber\\
&=&\rho_{\mathcal{K}},
\end{eqnarray}
where we used the fact that  $\mathrm{Tr}_{\mathcal{K}^c}\left[\rho_{\mathcal{J}}\right] = \mathrm{Tr}_{\mathcal{K}^c}\left[\sigma_{\mathcal{J}}\right] = \mathbb{I}_{\mathcal{K}^c}$.
\end{proof}
\noindent\textbf{Proposition 2} (Idempotence)
\textit{The marginal imposition operator $\mathcal{Q}_{\mathcal{J}}(\rho)$ is idempotent. That is, $\mathcal{Q}_{\mathcal{J}}\circ\mathcal{Q}_{\mathcal{J}}(\rho)=\mathcal{Q}_{\mathcal{J}}(\rho)$, for any subset of parties $\mathcal{J}$ and any marginal reduction $\sigma_{\mathcal{J}}$.}
\begin{proof}
From Definition 2, we have that \begin{eqnarray}
\mathcal{Q}_{\mathcal{J}}\circ\mathcal{Q}_{\mathcal{J}}(\rho)&=&\rho-\rho_{\mathcal{J}}+\sigma_{\mathcal{J}}-\mathrm{Tr}_{\mathcal{J}^c}(\rho-\rho_{\mathcal{J}}+\sigma_{\mathcal{J}})+\sigma_\mathcal{J}\nonumber\\
&=&\rho-\rho_{\mathcal{J}}+\sigma_{\mathcal{J}}\nonumber\\
&=&
\mathcal{Q}_{\sigma_{\mathcal{J}}}(\rho),
\end{eqnarray}
for any $\rho$ and $\sigma_\mathcal{J}$. Here, we recall that tensor product with maximally mixed reductions are omitted everywhere.

\end{proof}

\noindent\textbf{Proposition 3} (Closed form)
\textit{The composite marginal imposition operator is given by}
\begin{equation}
\mathcal{Q}_{\mathcal{J}_1,\dots,\mathcal{J}_m}(\rho)=\rho+\sum_{X\in\mathscr{J}}(-1)^{\nu(X)}(\rho_X-\sigma_X).
\end{equation}
\begin{proof}
The result is proven by induction in $m$. For $m=1$ we have a single marginal reduction, so $\mathscr{J}=\sigma_{J}$ and $\nu(X)=1$. This means that (\ref{qmo2}) reduces to Definition 2. Assuming that (\ref{qmo2}) holds for any $m>1$, let us show that it holds for $m+1$. Using (\ref{qmo2}) and Definition 2 with $J=J_{m+1}$, we have that
\begin{eqnarray*}
\mathcal{Q}_{\mathcal{J}_1,\dots,\mathcal{J}_{m+1}}(\rho)&=&\mathcal{Q}_{\mathcal{J}_{m+1}}\circ\mathcal{Q}_{\mathcal{J}_1,\dots,\mathcal{J}_{m}}(\rho)\\
&=&\mathcal{Q}_{\mathcal{J}_{m+1}}\left(\rho+\sum_{X\in\mathscr{J}}(-1)^{\nu(X)}(\rho_X-\sigma_X)\right)\\
&=&\rho+\sum_{X\in\mathscr{J}}(-1)^{\nu(X)}(\rho_X-\sigma_X)-\rho_{\mathcal{J}_{m+1}}-\sum_{X\in\mathscr{J}}(-1)^{\nu(X)}(\rho_{X\cap\mathcal{J}_{m+1}}-\sigma_{X\cap\mathcal{J}_{m+1}})+\sigma_{\mathcal{J}_{m+1}}\\
&=&\rho+\sum_{X\in\tilde{\mathscr{J}}}(-1)^{\nu(X)}(\rho_X-\sigma_X),
\end{eqnarray*}
where $\tilde{\mathscr{J}}$ is the set defined in (3), main text, with $m+1$. Therefore, equation (\ref{qmo2}) holds for any $m\in\mathbb{N}$.
\end{proof}

\noindent\textbf{Proposition 4} (Convexity)
\textit{Let $\rho$ and $\eta$ be any two multipartite quantum states acting on the same composite Hilbert space. Therefore, \begin{equation}
\mathcal{Q}_{\mathcal{J}_1,\dots,\mathcal{J}_m}(\lambda\rho+(1-\lambda)\eta)=\lambda\mathcal{Q}_{\mathcal{J}_1,\dots,\mathcal{J}_m}(\rho)+(1-\lambda)\mathcal{Q}_{\mathcal{J}_1,\dots,\mathcal{J}_m}(\eta),
\end{equation} 
for any $0\leq\lambda\leq1$, and any subsets of parties $\mathcal{J}_1,\dots,\mathcal{J}_m\subset\mathcal{I}$, not necessarily disjoined.}
\begin{proof}
From considering (\ref{qmo2}), and the linearity of the partial trace, we have that 
\begin{eqnarray*}
\mathcal{Q}_{\mathcal{J}_1,\dots,\mathcal{J}_m}(\lambda\rho+(1-\lambda)\eta)&=&(\lambda\rho+(1-\lambda)\eta)+\sum_{X\in\mathscr{J}}(-1)^{\nu(X)}\bigl((\lambda\rho+(1-\lambda)\eta)_X-\sigma_X\bigr)\\
&=&\lambda\rho+(1-\lambda)\eta+\lambda\sum_{X\in\mathscr{J}}(-1)^{\nu(X)}(\rho_X-\sigma_X)+(1-\lambda)\sum_{X\in\mathscr{J}}(-1)^{\nu(X)}(\eta_X-\sigma_X)\\
&=&\lambda\left(\rho+\sum_{X\in\mathscr{J}}(-1)^{\nu(X)}(\rho_X-\sigma_X)\right)+(1-\lambda)\left(\eta+\sum_{X\in\mathscr{J}}(-1)^{\nu(X)}(\eta_X-\sigma_X)\right)\\
&=&\lambda\mathcal{Q}_{\mathcal{J}_1,\dots,\mathcal{J}_m}(\rho)+(1-\lambda)\mathcal{Q}_{\mathcal{J}_1,\dots,\mathcal{J}_m}(\eta).
\end{eqnarray*} 
\end{proof}

\noindent\textbf{Proposition 5}
\textit{Two marginal imposition operators $\mathcal{Q}_{\mathcal{J}}$ and $\mathcal{Q}_{\mathcal{J}'}$ do commute if and only if the imposed marginals, $\sigma_{\mathcal{J}}$ and $\sigma_{\mathcal{J}'}$ respectively, are self-consistent.}
\begin{proof}
From Definition 2, we have that
\begin{equation*}
(\mathcal{Q}_{\mathcal{J}}\circ\mathcal{Q}_{\mathcal{J}'})(\rho)=\rho-\rho_{\mathcal{J}'}+\sigma_{\mathcal{J}'}-\bigl(\rho_{\mathcal{J}}-\mathrm{Tr}_{\mathcal{J}^c}(\rho_{\mathcal{J}'})+\mathrm{Tr}_{\mathcal{J}^c}(\sigma_{\mathcal{J}'})\bigr)+\sigma_{\mathcal{J}},
\end{equation*}
and
\begin{equation*}
(\mathcal{Q}_{\mathcal{J}'}\circ\mathcal{Q}_{\mathcal{J}})(\rho)=\rho-\rho_{\mathcal{J}}+\sigma_{\mathcal{J}}-\bigl(\rho_{\mathcal{J}'}-\mathrm{Tr}_{\mathcal{J}'^c}(\rho_{\mathcal{J}})+\mathrm{Tr}_{\mathcal{J}'^c}(\sigma_{\mathcal{J}})\bigr)+\sigma_{\mathcal{J}'}.
\end{equation*}
Taking into account the self-consistency relation $\mathrm{Tr}_{\mathcal{J}'^c}(\rho_{\mathcal{J}})=\mathrm{Tr}_{\mathcal{J}^c}(\rho_{\mathcal{J}'})$, we have that
\begin{equation*}
(\mathcal{Q}_{\mathcal{J}}\circ\mathcal{Q}_{\mathcal{J}'}-\mathcal{Q}_{\mathcal{J}'}\circ\mathcal{Q}_{\mathcal{J}})(\rho)=\mathrm{Tr}_{\mathcal{J}^c}(\sigma_{\mathcal{J}'})-\mathrm{Tr}_{\mathcal{J}'^c}(\sigma_{\mathcal{J}}).
\end{equation*}
Finally, the required commutativity arises from the fact that $\sigma_{\mathcal{J}}$ and $\sigma_{\mathcal{J}'}$ are self-consistent. That is, $$(\mathcal{Q}_{\mathcal{J}}\circ\mathcal{Q}_{\mathcal{J}'})(\rho)=(\mathcal{Q}_{\mathcal{J}'}\circ\mathcal{Q}_{\mathcal{J}})(\rho),$$ for any quantum state $\rho$, any pair of self-consistent reductions $\sigma_{\mathcal{J}}$ and $\sigma_{\mathcal{J}'}$, and any sets $\mathcal{J},\mathcal{J}'\subset\mathcal{I}$.
\end{proof}

\noindent\textbf{Proposition 6} (Fixed points)
\textit{An hermitian operator $\rho$ satisfies   $\mathrm{Tr}_{\mathcal{J}^c_1}(\rho)=\sigma_{\mathcal{J}_1},\dots,\mathrm{Tr}_{\mathcal{J}^c_m}(\rho)=\sigma_{\mathcal{J}_m}$ if and only if $\rho$ is a fixed point of the composite marginal imposition operator $\mathcal{Q}_{\mathcal{J}_1,\dots,\mathcal{J}_m}$.}
\begin{proof}
Suppose that $\mathrm{Tr}_{\mathcal{J}^c_1}(\rho)=\sigma_{\mathcal{J}_1},\dots,\mathrm{Tr}_{\mathcal{J}^c_m}(\rho)=\sigma_{\mathcal{J}_m}$. From Definition 2 and Proposition 1, the state $\rho$ is a fixed point of each operator  $\mathcal{Q}_{\mathcal{J}_1},\dots,\mathcal{Q}_{\mathcal{J}_m}$, associated to the imposition of marginals $\sigma_{\mathcal{J}_1},\dots,\sigma_{\mathcal{J}_m}$, respectively. Moreover, from Proposition 5, $\rho$ is a fixed point of the composite operator $\mathcal{Q}_{\mathcal{J}_1,\dots,\mathcal{J}_m}$, regardless of the ordering of the compositions. Reciprocally, suppose that $\rho$ is a fixed point of the composite operator $\mathcal{Q}_{\mathcal{J}_1,\dots,\mathcal{J}_m}$, for a given set of self-consistent marginal reductions $\sigma_{\mathcal{J}_1},\dots,\sigma_{\mathcal{J}_m}$. Therefore, from Proposition 1, the state $\rho$ contains the last imposed marginal reduction. As the marginal reductions arise from a generator state, they are compatible. Thus, from Proposition 5, the imposition operators $\mathcal{Q}_{\mathcal{J}_i}$ and $\mathcal{Q}_{\mathcal{J}_j}$ do commute, for any $i,j\in\{1,\dots,m\}$. As consequence, the state $\rho$ contains all marginal reductions $\sigma_{\mathcal{J}_j}$, $j=1,\dots,m$.
\end{proof}

\noindent\textbf{Proposition 7}
\textit{A set of marginal reductions $\sigma_{\mathcal{J}_1},\dots,\sigma_{\mathcal{J}_m}$ is compatible with the existence of a quantum state $\rho$ if and only if}
\begin{equation}
\sum_{X\in\mathscr{J}}(-1)^{\nu(X)}(\rho_X-\sigma_X)=0,\quad\mathrm{Tr}(\rho)=1,\quad\mbox{and}\quad\rho\geq0.
\end{equation}
\begin{proof}
From Proposition 6, a state $\rho$ contains the self-consistent marginal reductions $\sigma_{\mathcal{J}_1},\dots,\sigma_{\mathcal{J}_m}$ if and only if $\rho$ is a fixed point of $\mathcal{Q}_{\mathcal{J}_1,\dots,\mathcal{J}_m}$. On the other hand, Proposition 3 implies that
\begin{equation}
\mathcal{Q}_{\mathcal{J}_1,\dots,\mathcal{J}_m}(\rho)=\rho+\sum_{X\in\mathscr{J}}(-1)^{\nu(X)}(\rho_X-\sigma_X)=\rho,
\end{equation}
or, equivalently,
\begin{equation}
\sum_{X\in\mathscr{J}}(-1)^{\nu(X)}(\rho_X-\sigma_X)=0.
\end{equation}
The additional conditions $\mathrm{Tr}(\rho)=1$ and $\rho\geq0$ are required to produce a quantum state $\rho$.
\end{proof}

\noindent\textbf{Proposition 8}
\textit{Let $\sigma_{\mathcal{J}_1},\dots,\sigma_{\mathcal{J}_m}$ be a set of self-consistent marginal reductions. If $\mathcal{Q}_{\mathcal{J}_1,\dots,\mathcal{J}_m}(\rho)\geq0$, for a given quantum state $\rho$, then $\mathcal{Q}_{\mathcal{J}_1,\dots,\mathcal{J}_m}(\rho)$ is a quantum state containing all the above marginal reductions.}
\begin{proof}
Suppose that $\sigma_{\mathcal{J}_1},\dots,\sigma_{\mathcal{J}_m}$ is a set of self-consistent marginal reductions. From Proposition 5, all operators  $\mathcal{Q}_{\mathcal{J}_{\pi[1]},\dots,\mathcal{J}_{\pi[m]}}$ do commute, for any permutation of indices $\pi$. As a consequence of Proposition 1, $\sigma_{\mathcal{J}_1},\dots,\sigma_{\mathcal{J}_m}$ contains all marginal reductions $\sigma_{\mathcal{J}_1},\dots,\sigma_{\mathcal{J}_m}$. This is so because any operator $\mathcal{Q}_{\mathcal{J}_j}$ can be the last one applied, due to the commutation property. The proof concludes by taking into account that  $\mathcal{Q}_{\mathcal{J}_1,\dots,\mathcal{J}_m}(\rho)\geq0$, so it is a quantum state with the desired marginal reductions.  
\end{proof}

\noindent\textbf{Proposition 9}
\textit{Let $\sigma_{\mathcal{J}_1},\dots,\sigma_{\mathcal{J}_m}$ be a set of marginal reductions compatible with the existence of a quantum state $\sigma_{\mathcal{J}}$. Therefore, there is a set $V$ of quantum states $\rho$, having a positive volume in the Hilbert space, such that $\mathcal{Q}_{\mathcal{J}_1,\dots,\mathcal{J}_m}(\rho)\geq0$, for any $\rho\in V$. The result holds up to a null measure set of quantum states $\sigma_{\mathcal{J}}$.}
\begin{proof}
Suppose that $\sigma_{\mathcal{J}}$ is a quantum state having marginal reductions $\sigma_{\mathcal{J}_1},\dots,\sigma_{\mathcal{J}_m}$. Therefore, we have $\mathcal{Q}_{\mathcal{J}_1,\dots,\mathcal{J}_m}(\sigma_{\mathcal{J}})=\sigma_{\mathcal{J}}>0$, due to Proposition 6. The strict positivity here holds for all quantum states up to the set of rank $r<d$ quantum states $\sigma_{\mathcal{J}}$, which form a null measure set. This means that for the generic case there exists a ball $B_{\epsilon}(\sigma_{\mathcal{J}})$ with center at $\sigma_{\mathcal{J}}$ and radius $\epsilon$ such that $\mathcal{Q}_{\mathcal{J}_1,\dots,\mathcal{J}_m}(\sigma_{\mathcal{J}})=\sigma'_{\mathcal{J}}>0$, for any $\sigma'_{\mathcal{J}}\in B_{\epsilon}(\sigma_{\mathcal{J}})$ and a sufficiently small $\epsilon$.
\end{proof}

\noindent\textbf{Proposition 10}
\textit{Let $\sigma_{\mathcal{J}_1},\dots,\sigma_{\mathcal{J}_m}$ be any given set of self-consistent marginal reductions, according to Proposition 5. Therefore, the marginals $\tilde{\sigma}_{\mathcal{J}_j}=\Delta_{\epsilon}(\sigma_{\mathcal{J}_j})$, $j=1,\dots,m$, are compatible with the existence of a global quantum state, provided that $|\mathscr{J}|(1-\epsilon)(d-1)\leq1$, where $\mathscr{J}$ is defined in (3) [main text].}
\begin{proof}
The result arises from the fact that $\mathcal{Q}_{\mathcal{J}_1,\dots,\mathcal{J}_m}(\rho)>0$ when $\rho$ is the maximally mixed state and the imposed self-consistent marginal reductions are $\tilde{\sigma}_{\mathcal{J}_1},\dots,\tilde{\sigma}_{\mathcal{J}_m}$, for a sufficiently small value of $\epsilon$. Given that $\tilde{\sigma}_{\mathcal{J}_j}=\Delta_{\epsilon}(\sigma_{\mathcal{J}_j})=(1-\epsilon)\sigma_{\mathcal{J}_j}+\epsilon\mathbb{I}/d$, we have  
\begin{eqnarray}\label{Qepsilon}
\mathcal{Q}_{\mathcal{J}_1,\dots,\mathcal{J}_m}(\mathbb{I}/d)&=&\mathbb{I}/d+\sum_{x\in\mathscr{J}}(-1)^{\nu(X)}(\mathbb{I}/d-\tilde{\sigma}_X)\nonumber\\
&=&\mathbb{I}/d+(1-\epsilon)\sum_{x\in\mathscr{J}}(-1)^{\nu(X)}(\mathbb{I}/d-\sigma_X).
\end{eqnarray}
Equation (\ref{Qepsilon}) implies that the smallest eigenvalue of $\mathcal{Q}_{\mathcal{J}_1,\dots,\mathcal{J}_m}(\mathbb{I}/d)$, denoted $\lambda_{min}(\mathcal{Q}_{\mathcal{J}_1,\dots,\mathcal{J}_m}(\mathbb{I}/d))$, satisfies
\begin{eqnarray}\label{lbound}
\lambda_{min}(\mathcal{Q}_{\mathcal{J}_1,\dots,\mathcal{J}_m}(\mathbb{I}/d))&\geq&\frac{1}{d}+(1-\epsilon)|\mathscr{J}|\min_X \lambda_{min}(\mathbb{I}/d-\sigma_X)\nonumber\\
&\geq&\frac{1}{d}+(1-\epsilon)|\mathscr{J}|\frac{1-d}{d}\nonumber\\
&\geq&\frac{1-|\mathscr{J}|(1-\epsilon)(d-1)}{d}.
\end{eqnarray}
Note that the first inequality in (\ref{lbound}) holds because $\min_X \lambda_{min}(\mathbb{I}/d-\sigma_X)<0$, regardless on the set of marginals $\{\sigma_X\}$. We also used the basic fact that given two hermitian operators $A$ and $B$, it holds that $\lambda_{min}(A+B)\geq\lambda_{min}(A)+\lambda_{min}(B)$. Thus, we obtain from (\ref{lbound}) that $\mathcal{Q}_{\mathcal{J}_1,\dots,\mathcal{J}_m}(\mathbb{I}/d)\geq0$, whenever $|\mathscr{J}|(1-\epsilon)(d-1)\leq1$. Finally, the marginal reductions $\tilde{\sigma}_{\mathcal{J}_1},\dots,\tilde{\sigma}_{\mathcal{J}_m}$ are self-consistent, implying that the positive semidefinite -and normalized- operator $\mathcal{Q}_{\mathcal{J}_1,\dots,\mathcal{J}_m}(\mathbb{I}/d)$ contains all these marginals, due to Proposition 5.  
\end{proof}

\subsection*{Supplemental Material B: Analytical expressions}
\noindent\textbf{Analytical expression (8):} \textit{Let $\mathcal{I}=\{ A_0, A_1, \ldots, A_{N-1} \}$. Therefore, we have}
\begin{equation}\label{QN2k1}
   \mathcal{Q}_{A_{N-1}} \circ \ldots \circ \mathcal{Q}_{A_0}(\mathbb{I})= \underset{ i \in \mathcal{I} }{ \sum } \sigma_i -(N-1)\mathbb{I}.
\end{equation}
\begin{proof}
Straightforward by taking partial trace on $\mathcal{Q}_{A_{N-1}} \circ \ldots \circ \mathcal{Q}_{A_0}(\mathbb{I})$ over all parties except $A_i$, thus obtaining $\sigma_{A_i}$. \end{proof}
\noindent\textbf{Analytical expression (9):} \textit{Let $\mathcal{J}=\{ A_0A_1,A_0A_2\dots,A_0A_{N-1},A_1A_2,A_1A_3,\dots,A_{N-2}A_{N-1}\}$ be  the set of all bipartite labels and $\mathcal{I}=\{ A_0, A_1, \ldots, A_{N-1} \}$. Therefore, we have}
\begin{equation}\label{QN2}
   \mathcal{Q}_{A_{N-2}A_{N-1}} \circ \ldots \circ \mathcal{Q}_{A_0A_1}(\mathbb{I})= \underset{  j \in \mathcal{J}  }{ \sum } \sigma_j - (N-2)\underset{ i \in \mathcal{I} }{ \sum } \sigma_i +\left( 1+\frac{N^2}{2}-\frac{3}{2}N\right)\mathbb{I}.
\end{equation}
\begin{proof}
The proof is based on taking a reduction of the quantum state (\ref{QN2}) with respect to the bipartite subsystem $\{A_iA_j\}\subset\mathcal{J}$. Let us explain term by term the reduction. First,
\begin{equation}\label{QN2_2}
\left(\underset{  j \in \mathcal{J}  }{ \sum } \sigma_j\right)_{A_iA_j}=\sigma_{A_iA_j}+(N-2)(\sigma_{A_i}+\sigma_{A_j})+\left(\frac{N(N-1)}{2}-1-2(N-2)\right)\mathbb{I},
\end{equation}
where the second term is associated to the appearance of $N-2$ times each single party marginal $\sigma_{A_i}$ and $\sigma_{A_j}$. The third term corresponds to all cases where a parties $A_i$ and $A_j$ does not occur, so having the maximally mixed state. Next, we take a bipartition over the sum of all the single party marginal reductions, 
\begin{equation}\label{QN2_3}
\left(\underset{ i \in \mathcal{I} }{ \sum } \sigma_i\right)_{A_iA_j}=\sigma_{A_i}+\sigma_{A_j}+(N-2)\mathbb{I},
\end{equation}
where, again, the last term corresponds to partial traces from cases where $A_i$ and $A_j$ do not occur.

From considering the reduction of (\ref{QN2}) to parties $A_i$ and $A_j$, and taking into account (\ref{QN2_2}) and (\ref{QN2_3}), we have that
\begin{equation}
\left(\mathcal{Q}_{A_{N-2}A_{N-1}} \circ \ldots \circ \mathcal{Q}_{A_0A_1}(\mathbb{I})\right)_{A_iA_j}=\sigma_{A_iA_j}.
\end{equation}
\end{proof}
Similarly, further expressions can be found for a higher number of reductions. For instance, for $N=5$ and all reductions to $k=3$ parties we have
\begin{eqnarray}
\label{N5k3}
\mathcal{Q}_{CDE} \circ \ldots \circ \mathcal{Q}_{ABC}(\mathbb{I}) &=& \left( \sigma^{ABC} + \sigma^{ABD} + \sigma^{ABE} + \sigma^{ACD} + \sigma^{ACE} + \sigma^{ADE} + \sigma^{BCD} + \sigma^{BCE}\right.\nonumber\\
&& \left.+ \sigma^{BDE} + \sigma^{CDE} \right)  - 2\left( \sigma^{AB} + \sigma^{AC} + \sigma^{AD} + + \sigma^{AE} + \sigma^{BC} + \sigma^{BD}\right.\nonumber \\
&&\left. + \sigma^{BE} + \sigma^{CD} + \sigma^{CE} + \sigma^{DE} \right) + 3\left( \sigma^A + \sigma^B + \sigma^C + \sigma^D  + \sigma^E\right) - 4\mathbb{I}.\nonumber\\
\end{eqnarray}
It might be the case that a general formula can be derived for any number of parties $N$ and any number of reductions $k<N$. However, we were not able to find it.

\subsection*{Supplemental Material C: Iterative algorithm}\label{sec:alg}
Proposition 10 shows a way to find a global quantum state from any given set of self-consistent quantum marginal reductions that are sufficiently close to the maximally mixed state. However, there are many interesting physical situations where this condition does not hold. For instance, this result cannot be used when considering the 2-body reductions of the 3-partite $W$ state. Despite a general analytical solution is hopeless, we define an algorithm to numerically solve the problem, based on dynamical systems theory and the physical imposition operator.

As we have shown in Proposition 1, the image of the composite operator $\mathcal{Q}_{\mathcal{J}_1,\dots,\mathcal{J}_m}$ already contains all the marginal reductions $\sigma_{\mathcal{J}_1},\dots,\sigma_{\mathcal{J}_m}$. However, this is not a solution to the quantum marginal problem because the image of $\rho$ is not necessarily positive semidefinite. In order to find a compatible quantum state, when it exists, we introduce an iterative algorithm that alternates between imposing marginal reductions and projecting over the set of quantum states. From now on, $\mathcal{P}_r(\rho)$ denotes the projection of an hermitian operator over the subspace generated by the eigenvectors of $\rho$ associated to its $r$ largest eigenvalues $\lambda_{0} \geq \lambda_{1} \ldots \geq\lambda_{r-1}$, rescaled so that $\mathrm{Tr}[\mathcal{P}_r(\rho)]=1$. The algorithm runs until the total distance $\mathcal{D}_{T} = \sqrt{\mathcal{D}_{\lambda}^2 + \mathcal{D}_{M}^2} < \epsilon$, for a predefined $\epsilon$, where $\mathcal{D}_{\lambda}= \sqrt{\sum_{i=r}^d|\lambda_{i}|^2}$  and  $ \mathcal{D}_{M} = \sqrt{ m^{-1} \sum_{i=1}^m \mathfrak{D}(\sigma_{\mathcal{J}_i},\sigma^n_{\mathcal{J}_i})^2}$. Here, $\mathfrak{D}(\sigma_{\mathcal{J}_i},\sigma^n_{\mathcal{J}_i}):=\mathrm{Tr}[(\sigma_{\mathcal{J}_i}-\sigma^n_{\mathcal{J}_i})^2]$ denotes the Hilbert-Schmidt distance between the imposed marginal $\sigma_{\mathcal{J}_i}$ and the resulting marginal $\sigma^n_{\mathcal{J}_i}$, after $n$ iterations. The algorithm is defined as follows:
\begin{figure}[h!]
        \subfloat[\label{fig:ame52a}]{%
       \includegraphics[scale = 0.36]{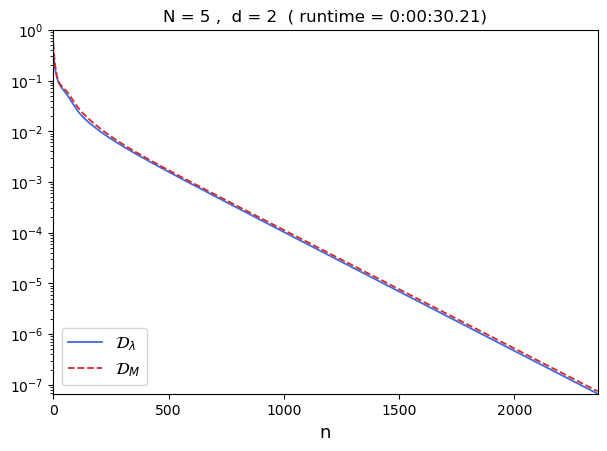}}         \subfloat[\label{fig:ame62a}]{%
       \includegraphics[scale = 0.36]{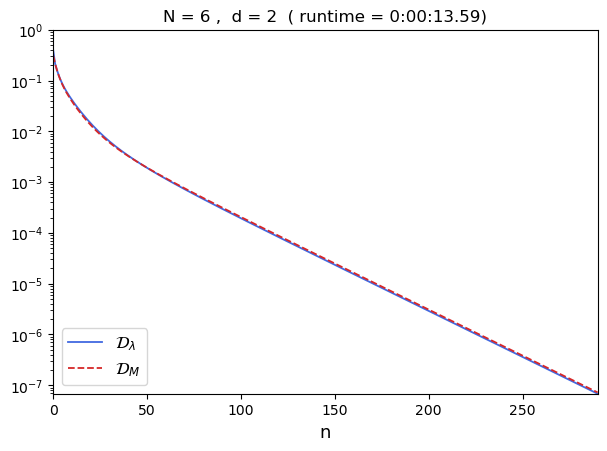}}
     \subfloat[\label{fig:ame43a}]{%
       \includegraphics[scale = 0.36]{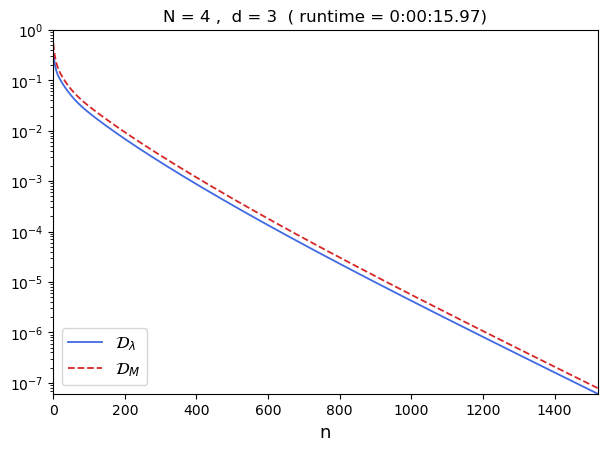}}\\
     \subfloat[\label{fig:ame52b}]{%
       \includegraphics[scale = 0.36]{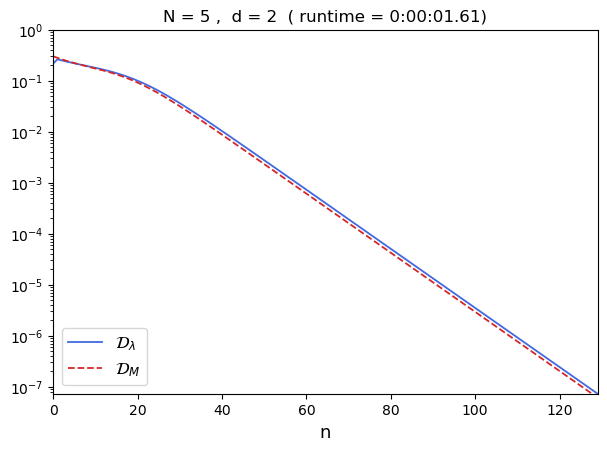}}
    \subfloat[\label{fig:ame62b}]{%
       \includegraphics[scale = 0.36]{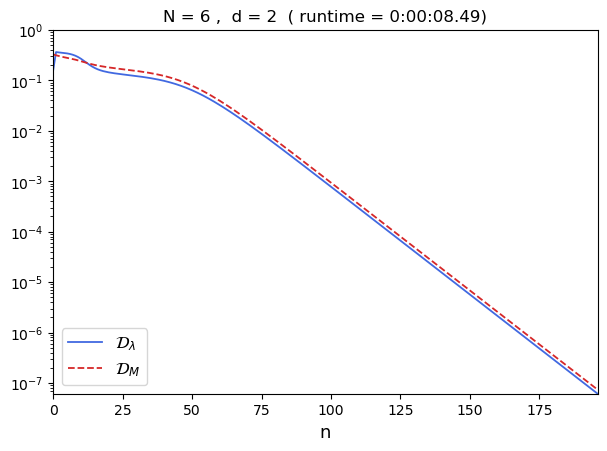}}
     \subfloat[\label{fig:ame43b}]{%
       \includegraphics[scale = 0.36]{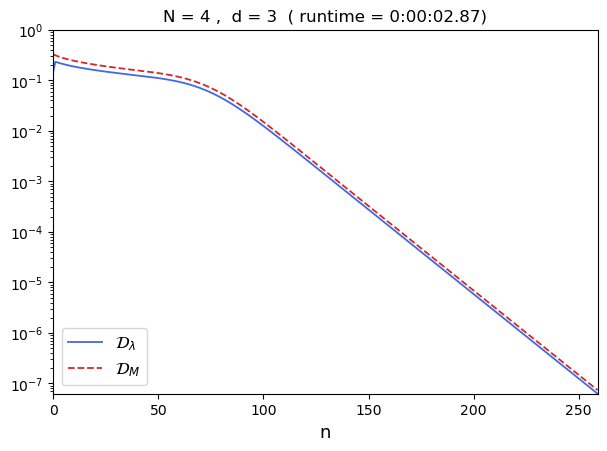}}
     \caption{Convergence of Algorithm \ref{alg:qmp} for global pure states, in the case of $N=4,5,6$ parties. Here, we impose all reductions to $\lfloor N/2\rfloor$ parties, where $\lfloor\cdot\rfloor$ denotes the floor function. In figures (\ref{fig:ame52a}), (\ref{fig:ame62a}) and (\ref{fig:ame43a}), the marginal reductions arise from from a pure generator state $\rho_{gen}$, chosen at random according to the Hilbert-Schmidt measure. On the other hand, in Figures (\ref{fig:ame52b}), (\ref{fig:ame62b}) and (\ref{fig:ame43b}), we impose maximally mixed reductions to $\lfloor N/2\rfloor$ parties, thus obtaining AME states. The vertical axis corresponds to $\mathcal{D}_{\lambda}$ (solid blue) and $\mathcal{D}_{M}$ (dashed red) in logarithmic scales, whereas the horizontal axis shows the number of iterations of the algorithm, denoted by $n$. Note that both distances monotonically tend to zero in all the cases, until achieving accuracy $\epsilon=10^{-7}$. }
     \label{fig1}
\end{figure}

\begin{algorithm}\caption{Quantum marginal problem}\label{alg:qmp}
\begin{algorithmic}
\Require Dimension $d\in\mathbb{N}$, self-consistent set of marginal reductions $\sigma_{\mathcal{J}_1},\dots,\sigma_{\mathcal{J}_m}$, and accuracy $\epsilon\in [0,1]$.
\Ensure estimate $\rho_{\rm est}\in\mathcal{B}(\mathcal{H}_d)$.
\State{$\rho_{0}$ (randomly chosen)}
\Repeat{\\
\hspace*{0.5cm}$\rho_1 = \mathcal{Q}_{\mathcal{J}_1,\dots,\mathcal{J}_m}(\rho_{0})$}\\
\hspace*{0.5cm}$\rho_0 = \mathcal{P}_r(\rho_1)$
\Until{ $\mathcal{D}_{T} < \epsilon$ }
\end{algorithmic}
\end{algorithm}
In Figure \ref{fig1}, we show the behaviour of Algorithm \ref{alg:qmp} in the case of $N=4,5,6$ parties. Quantum states are reconstructed from all $\lfloor N\rfloor$ marginal reductions, defined as follows: \textit{1a), 1b) and 1c)} marginals generated from a pure quantum state defined at random, according to the Haar measure \textit{1d) 1e) 1f)} maximally mixed reductions. This last case allowed us to recover the AME(5,2), AME(6,2) and AME(4,3) states, respectively, with a small runtime. Despite this success, we were not able to conclude that the AME(6,4) state exists, which has been confirmed recently [see Phys. Rev. Lett. \textbf{128}, 8, 080507 (2022)].

\subsection*{Supplemental Material D: Fraction of positivity}
In this section, we study the fraction of randomly chosen generator states $\sigma$ for which the operator  $\mathcal{Q}_{\mathcal{J}_1,\dots,\mathcal{J}_m}(\mathbb{I}/d)$ is positive semidefinite. In other words, this study shows the fraction of the space of compatible marginal reductions that allows a state reconstruction through $\mathcal{Q}_{\mathcal{J}_1,\dots,\mathcal{J}_m}(\mathbb{I}/d)$. In Table I, we consider the reconstruction of $n$-qubit quantum states from all its $k$-body marginal reductions, which are estimated from an $n$-qubit generator state $\sigma$, for $n\leq8$. Here, we are not interested to determine whether the reconstruction is unique or not. We considered 1000 generator states $\sigma$ to estimate the fractions when $3\leq n\leq 8$. As the runtime excessively increases with the number of qubits $n$, we considered 100 generators for $n=9$, and  10 generators when $n\geq10$. Despite that, it is remarkable that we obtain a fraction equal to 1 in some scenarios, even for a small number of randomly chosen generators. 

It is important to emphasize that this study does not involve the implementation of any multi-step algorithm but the consideration of the closed form given in Proposition 4, with a maximally mixed seed. Remarkably, the cases highlighted in gray in Table I exhibit a successful reconstruction for any randomly chosen generator state $\sigma$. This evidence strongly indicates that we can produce a global quantum state $\mathcal{Q}_{\mathcal{J}_1,\dots,\mathcal{J}_m}(\mathbb{I}/d)$ from almost any set of marginal reductions $\sigma_{\mathcal{J}_1},\dots,\sigma_{\mathcal{J}_m}$, compatible with the existence of a quantum state.

\begin{table}
\begin{tabular}{|c|c|c|c|c|c|c|c|c|c|c|}
\hline
\backslashbox{$k$}{$n$}&3&4&5&6&7&8&9&10&11&12\\
\hline
2&0.667& 0.998 & \colorbox{gray!20}{1.000} & \colorbox{gray!20}{1.000}&\colorbox{gray!20}{1.000}&\colorbox{gray!20}{1.000}&\colorbox{gray!20}{1.000}&\colorbox{gray!20}{1.000}&\colorbox{gray!20}{1.000}&\colorbox{gray!20}{1.000}\\
\hline
3& - & 0.001 & 0.479 & \colorbox{gray!20}{1.000}&\colorbox{gray!20}{1.000}&\colorbox{gray!20}{1.000}&\colorbox{gray!20}{1.000}&\colorbox{gray!20}{1.000}&\colorbox{gray!20}{1.000}&\colorbox{gray!20}{1.000}\\
\hline
4&-&-& 0.000 & 0.000 &\colorbox{gray!20}{1.000}&\colorbox{gray!20}{1.000}&\colorbox{gray!20}{1.000}&\colorbox{gray!20}{1.000}&\colorbox{gray!20}{1.000}&\colorbox{gray!20}{1.000}\\
\hline
5&-&-&-& 0.000&0.000&0.044&\colorbox{gray!20}{1.000}&\colorbox{gray!20}{1.000}&\colorbox{gray!20}{1.000}&\colorbox{gray!20}{1.000}\\
\hline
6&-&-&-&-&0.000&0.000&0.000&\colorbox{gray!20}{1.000}&\colorbox{gray!20}{1.000}&\colorbox{gray!20}{1.000}\\
\hline
7&-&-&-&-&-&0.000&0.000&0.000&\colorbox{gray!20}{1.000}&\colorbox{gray!20}{1.000}\\
\hline
8&-&-&-&-&-&-&0.000&0.000&0.000&0.000\\
\hline
9&-&-&-&-&-&-&-&0.000&0.000&0.000\\
\hline
10&-&-&-&-&-&-&-&-&0.000&0.000\\
\hline
11&-&-&-&-&-&-&-&-&-&0.000\\
\hline
\end{tabular}\hspace{2cm}
\label{tabla1}
\caption{ Fraction of cases where $\mathcal{Q}_{\mathcal{J}_1,\dots,\mathcal{J}_m}(\mathbb{I}/d)$ is a quantum state containing a given set of self-consistent marginal reductions $\sigma_{\mathcal{J}_1},\dots,\sigma_{\mathcal{J}_m}$, defined through a generator mixed state $\sigma_{\mathcal{J}}$, chosen at random according to the Hilbert-Schmidt measure. Note that for $\lfloor(n-1)/\sqrt{2}\rfloor\leq k\leq n-1$  and $5\leq n\leq12$, we have the fraction equal to 1, where $\lfloor\cdot\rfloor$ denotes the floor function.} 
\end{table}




















\bibliography{references}

\end{document}